\documentclass[12pt]{iopart}
\usepackage{graphicx,dcolumn,bm,mathptmx,etoolbox,ulem,xcolor}

\usepackage[utf8]{inputenc}
\usepackage[T1]{fontenc}

\begin{document}
\title{An in-situ thermoelectric measurement apparatus inside a thermal-evaporator}

\author{Kien Trung Nguyen}
\address{Nano and Energy Center, VNU University of Science, Vietnam National University, 120401 Hanoi, Vietnam}
\address{Faculty of Physics, VNU University of Science, Vietnam National University, 120401 Hanoi, Vietnam}
\author{Giang Bui-Thanh}
\address{Department of Advanced Materials Science and Nanotechnology, University of Science and Technology of Hanoi, Hanoi, Viet Nam}
\author{Hong Thi Pham}
\address{Nano and Energy Center, VNU University of Science, Vietnam National University, 120401 Hanoi, Vietnam}
\address{Faculty of Physics, VNU University of Science, Vietnam National University, 120401 Hanoi, Vietnam}
\author{Thuat Nguyen-Tran}
\address{Nano and Energy Center, VNU University of Science, Vietnam National University, 120401 Hanoi, Vietnam}
\author{Chi Hieu Hoang}
\address{Faculty of Physics, VNU University of Science, Vietnam National University, 120401 Hanoi, Vietnam}
\author{Hung Quoc Nguyen}
\ead{hungnq@hus.edu.vn}
\address{Nano and Energy Center, VNU University of Science, Vietnam National University, 120401 Hanoi, Vietnam}

\date{\today}

\begin{abstract}
At the ultra-thin limit below 20 nm, a film's electrical conductivity, thermal conductivity, or thermoelectricity depends heavily on its thickness. In most studies, each sample is fabricated one at a time, potentially leading to considerable uncertainty in later characterizations. We design and build an in-situ apparatus to measure thermoelectricity during their deposition inside a thermal evaporator. A temperature difference of up to 2 K is generated by a current passing through an on-chip resistor patterned using photolithography. The Seebeck voltage is measured on a Hall bar structure of a film deposited through a shadow mask. The measurement system is calibrated carefully before loading into the thermal evaporator. This in-situ thermoelectricity measurement system has been thoroughly tested on various materials, including Bi, Te, and Bi$_2$Te$_3$, at high temperatures up to 500 K. Working reliably and precisely, the in-situ measurement system would help to study physics during film growth or speedup our search for better themoelectric materials. 
\end{abstract}


\section{Introduction}

Seebeck coefficient is an essential parameter for the design of thermoelectric devices, such as thermoelectric generators and coolers. Quantum confinement effects can significantly alter the Seebeck coefficient of a material, making it a key factor to consider when designing thermoelectric devices. Quantum confinement occurs when the size of a material is reduced to the nanoscale, resulting in a change in its electronic structure. Depending on the material and the confinement size, the Seebeck coefficient can increase or decrease. There is much theoretical and experimental research on the Seebeck coefficient of thin film materials. These can be mentioned as the Seebeck coefficient of the material is enhanced when the size is reduced to one-dimensional \cite{Hicks1D}, two-dimensional \cite{Hicks2D} materials, and the influence of polycrystalline structures on Seebeck coefficient in the ultra-thin thickness region \cite{Bach}. 

However, the Seebeck coefficient fluctuates when the film reaches a few atomic layers thickness due to the emergence of subbands near the Fermi level \cite{Freik}. By increasing carrier concentration, quantum conﬁnement also improves the electrical conductivity $\sigma$ \cite{Ren, Kanatzidis}. Overall, the ﬁgure of merit $ZT = S^2\sigma/\kappa$, characterizes the qualities of thermoelectric materials and fluctuates sharply with film thickness \cite{Hicks1D, Hicks2D, Bach, Venkatasubramanian}. Here, $\kappa$ denotes the thermal conductivity. The prediction of those theoretical models has been verified by several experimental groups, for example, PbTe or PbEuTe thin films grown by molecular-beam epitaxy \cite{Hicks96} or Bi$_2$Te$_3$ thin film fabricated by thermal co-evaporation \cite{Thao}. In these works, different samples are optimized by ex-situ measurement of the thermoelectric coefficient. Samples are independently fabricated with different thicknesses and other parameters and measured after the fabrication. Moreover, the Seebeck coefficient is also found to fluctuate at the ultra-thin regime. Specifically, with Bi$_2$Te$_3$ thin film fabricated on glass substrates using thermal evaporation, the oscillation of the Seebeck coefficient depends on thickness below 100 nm. It becomes stable similar to bulk material value above 100 nm \cite{Rogacheva9, Rogacheva10, Rogacheva11}. This result is also reproduced on BiSe thin films fabricated using thermal evaporation on a glass substrate \cite{Rogacheva12}. Similar results with chalcogenide materials such as PbS, PbSe, PbTe \cite{Rogacheva13}, or SnSe \cite{Rogacheva14} thin films on KCl/EuS substrates recorded the combination of fluctuations and the increase of the thickness-dependent Seebeck coefficient when the thin thickness is below 100 nm. 

All these studies share a common feature: each sample is made one by one, followed by sophisticated measurements \cite{Venkatasubramanian, Hicks1D,Rogacheva9, Rogacheva10, Rogacheva11, Rogacheva12, Rogacheva13, Rogacheva14, DuongAT, PhanBT}. The properties of each sample depend on many fabrication details, such as the substrate temperature, its surface cleanliness, the stoichiometry of each component, vacuum chamber pressure, or cross-contaminations. Although these parameters are well controlled and calibrated, random fluctuations are always present, which lead to variations between different batches, especially at the ultra-thin film limit when transport exponentially depends on thickness. The polycrystal film consists of multiple grains, and it is hard to precisely determine a thickness different to a tenth of a nanometer. Moreover, each run might result in a slightly different film morphology due to random factors in the preparation steps. Because of these limitations, we propose measuring thin film thermoelectric properties during their deposition. This way, all data are measured from one experiment, avoiding random fluctuations. Instead of making many samples at various times, our in-situ measurement setup will shorten the experiment time and directly probe the thermoelectric properties during film formation. More importantly, data obtained from a single sample would be unique and more reliable. The in-situ measurement approach might have some limitations, such as the structure of the sample can change after the deposition. It might not be straightforward to conclude that the in-situ measurement result is accurate and replaces the ex-situ traditional measurement method. However, the development of the in-situ measurement is essential to observe the smallest changes in the sample as the thickness is in the ultrathin region where the quantum confinement and the effects of grains morphology strongly effect to all thermoelectric properties of materials, including the Seebeck coefficient, the electrical conductivity, and the thermal conductivity. It might also improve our process in optimizing thermoelectric materials. To our knowledge, nobody have measured thermoelectric properties of thin films during their thermal co-evaporation.

In this work, we design and construct an in-situ measurement setup inside a thermal co-evaporator. Although it is built to work for all proper materials, we focus our study on the thermoelectricity of Bi$_2$Te$_3$ thin films, a popular material with excellent thermoelectric properties at room temperature. The ﬁgure of merit of Bi$_2$Te$_3$ is high at room temperature \cite{Gonzalez16}, leading to its popularity, especially in thermoelectric cooling systems. There are many methods that can be used to fabricate Bi$_2$Te$_3$ thin films like exfoliation \cite{Balandin17}, molecular-beam epitaxy \cite{Hicks2D}, magnetron sputtering \cite{Shinohara18}, pulse laser deposition \cite{Leu19}, chemical vapor deposition \cite{Jitov20}, or thermal evaporation. In the thermal evaporation method, there are numerous variations such as evaporation from one source \cite{Fedorov21}, two-step single-source thermal evaporation \cite{Zheng22}, co-evaporation \cite{Min23,Correia24,Dang25}, or combined with annealing \cite{Oh26}. In this work, we co-evaporate Bi$_2$Te$_3$ from two sources, which are known to produce high-quality thermoelectric films \cite{Thao}.

\section{Method}

\begin{figure}[t]
\begin{center}
\includegraphics[width=3in,keepaspectratio]{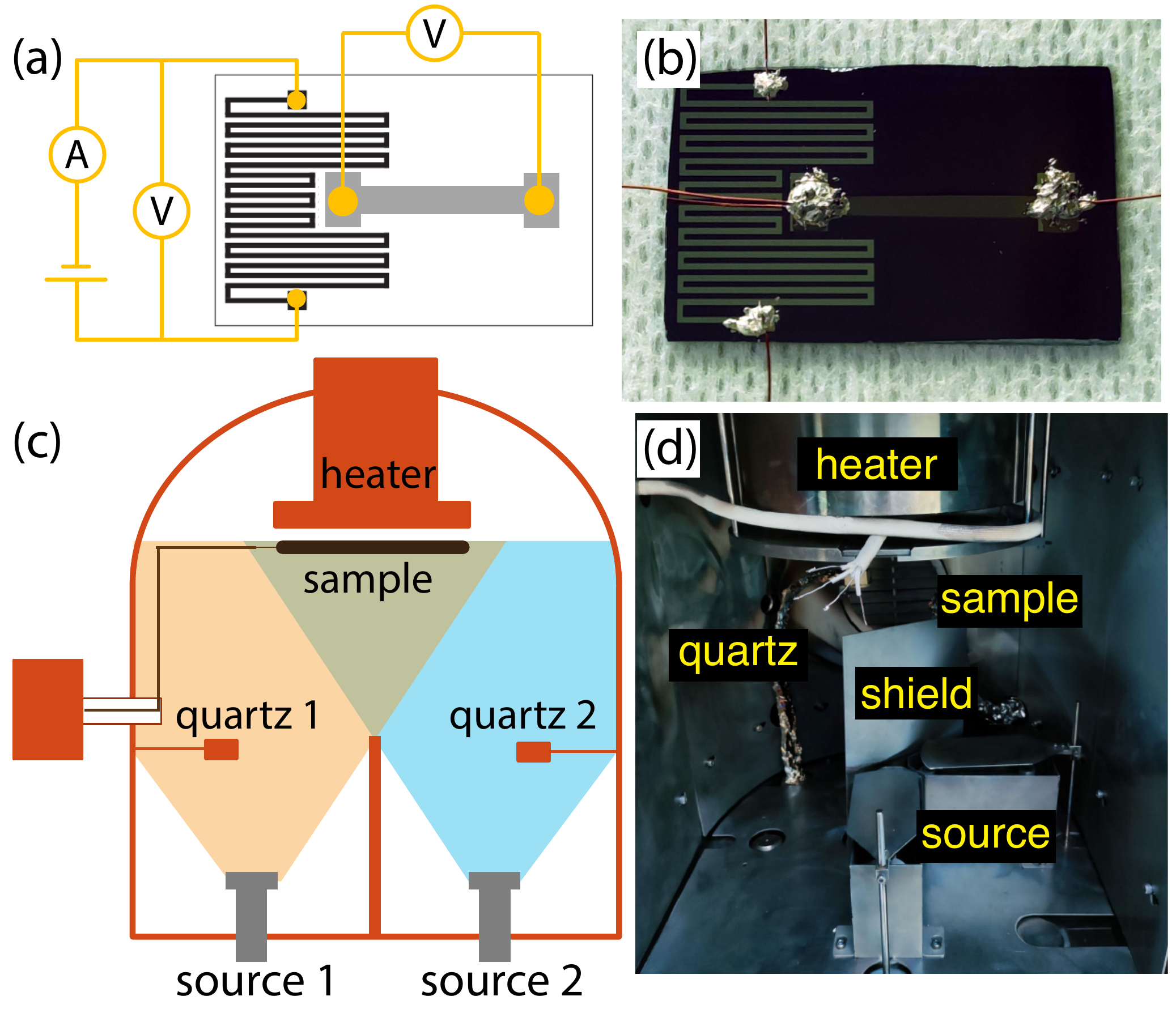}
\caption{\textbf{The in-situ thermoelectric measurement system:} (a) Layout of the on-chip heater circuit diagram with a meander shape metallic wire. Its width is 0.4 mm, while the total length is 240 mm, which constitutes 600 square sheet resistance. A thin silver film is deposited and patterned with a lift-off process. Duralco 124 silver paint glues Cu wires of 0.5 mm diameter to the metallic pads. This chip is covered with a metallic shadow mask of 12 $\times$ 2 mm bar shown in gray color. One end of the bar is surrounded by the on-chip heater, which generates an elevated temperature of a few Kelvin, depending on the applied current. The other end of the bar is the cold end and has no heater. (b) Optical micrograph of the actual Bi$_2$Te$_3$ sample after evaporation. The silver paint blobs are visible in grayish color. (c) Diagram of the modified vacuum chamber. With a volume of 125000 cm$^3$, the sample is 40 cm away from the two sources. To prevent mixing, a metallic shield is placed between the sources. Two Inficon quartz crystals monitor flow rates at 5 MHz excitations. The sample is placed in the deposit region through a shadow mask with four copper wires attached to the contact pad of the sample. These wires are then connected to electronics outside the chamber for automated data logging. (d) Image of the modified chamber of the thermal evaporator.}
\label{fig1}
\end{center}
\end{figure}

As shown in Fig.~\ref{fig1}, the in situ measurement apparatus is built to fit in a thermal evaporator used to study thermoelectricity in Bi$_2$Te$_3$. It must fit inside a vacuum system and work reliably at high temperatures. Most parts are made of stainless steel to maintain a high vacuum and avoid degassing. A Syskey model TH-01 thermal evaporator is modified to meet the requirement for the co-evaporation process. This is a standard three-source evaporator with a chamber volume of 125000 cm$^3$ and a source-sample distance of 40 cm. A metallic sheet is installed between the two sources to prevent the mixing of materials at the sources and the quartz sensors during evaporation. Two Inficon quartz crystals are installed near these sources so that the deposition rates of the two materials can be monitored independently using 5 MHz excitations. 

To connect the sample to the measurement system, copper wires of 0.5 mm diameter with insulating cover are glued to conducting pads, see Fig. 1b. A special silver glue that withstands high temperatures up to 500 K during Bi$_2$Te$_3$ deposition is compulsory. Duralco 124 silver ﬁlled epoxy, which has high chemical resistance and stability at temperatures up to 600 K, is selected to contact Cu wires to the film's contact pads. According to its manual, this epoxy cures at 393 K for 4 hours and post-cures at 453 K for 4 hours to have good contacts and to be applied for further use. The process of attaching silver epoxy is complicated and must be done with care. Applying a nice and thin layer of this silver epoxy is important, as rough steps block the connectivity between the ultra-thin film and the conducting paste. After attaching silver epoxy to the contact pads, the copper wire ends will be soldered to the homemade feedthrough system to connect to the external measuring equipment. Here, all connections are BNC type combined with coaxial cables grounded carefully to reduce noise. The wires are housed inside a double-shielded metal box connected to an epoxy-filled KF-40 flange feedthrough. All cable pairs are twisted to reduce cross talks. The evaporator is designed with a rotating motor to increase the uniformity of the sample. Due to the complicated wiring of the experiment, the substrate does not rotate during evaporations. Although we never noticed any difference, this lack of rotation of the base could reduce the film homogeneity due to the short source-sample distance in the evaporator. 

\begin{figure}[t]
\begin{center}
\includegraphics[width=3in,keepaspectratio]{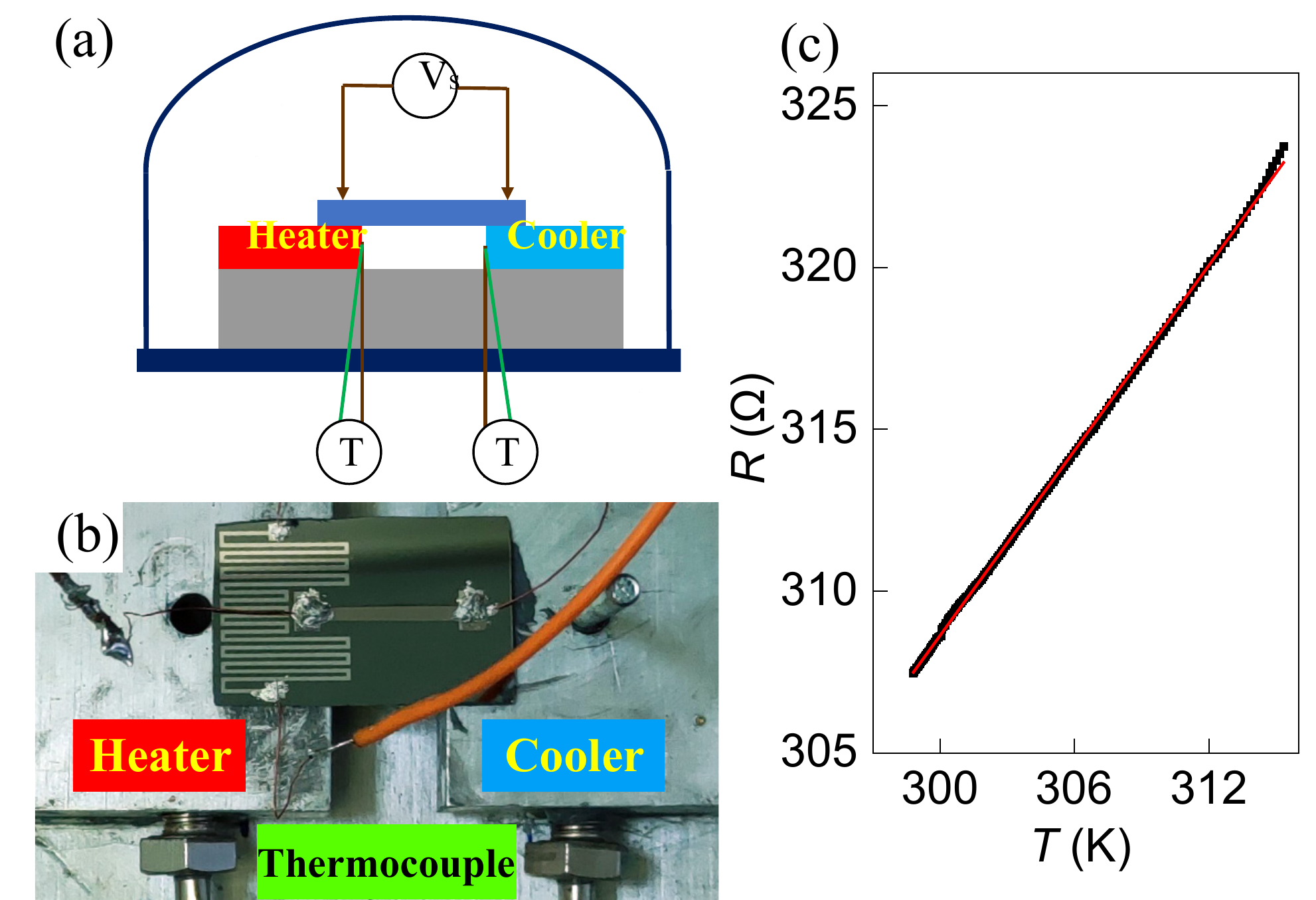}
\caption{\textbf{The ex-situ thermoelectric measurement system: }(a) Diagram of the measurement setup: a temperature gradient is provided homogeneously by two copper blocks, whose temperatures are precisely controlled using two commercial thermoelectric coolers. To reduce noise and prevent convection, the setup is housed inside a metallic vacuum chamber with help from a computer and standard electronics. (b) Real image of the sample being mounted for calibration. The electrodes are attached to the wires using standard silver paint, and the wires are connected to multimeters via BNC connections. The whole system is placed in a vacuum chamber to prevent electrical noises and thermal convection. (c) The dependence of the resistance on the temperature of the on-chip heater. Its slope yields the temperature coefficient of resistance and approximates the bulk material's value. Here, the resistance value is determined by the formula $R$ = 282.4 + 0.99$\Delta T$ ($\Omega$), which gives 3.5$\times$10$^{-3}$ K$^{-1}$. This value is quite close to that of bulk silver. }
\label{fig2}
\end{center}
\end{figure}

Before the main experiment, an on-chip heater is fabricated using lithography and lift-off. A silicon chip with size 3 cm $\times$ 2 cm is cleaned in Piranha acidic solution for 2 hours at 400 K, followed by a 10 minutes dip in acetone and isopropanol, and then nitrogen blows dry. A thin layer of AZ ECI resist is spun coat at 4000 rpm and baked at 373 K for 1 minute. A 36 mJ/cm$^2$ exposure in the I-line of the mercury-vapor lamp followed by 30-second develop in 2.5\% TMAH solution finishes the negative lithography step. Here, a meander pattern of width 0.4 mm and total length 240 mm, which constitute a 600 square sheet, is fabricated. The chip is then pumped down in the evaporator, and a 100 nm thickness of metal is thermally deposited. In most cases, silver is used due to its apparent practicality. A lift-off in acetone removes the resist and completes the on-chip heater. Typically, the resistance of the heater is in the range of 200 $\Omega$. We do not notice any difference between the type of metal, say Au, Ag, Cu, Al, or Cr. At least 3 to 5 chips are fabricated per run to speed up the process.

To measure this film, a thin layer of silver paint glues thin Cu wires to the pads. It is important to fix a small and thin paste such that even a few nm films conduct. Silver epoxy is manually applied to connect the 0.5 mm diameter copper wire to the conducting pads of the sample. To survive the high temperature up to 250 Celsius required by some materials, for example, Bi$_2$Te$_3$, the silver paint must be of a particular type. We chose Duralco 124 silver paint for such a purpose. The sample is baked for 8 hours as required by the silver epoxy. When the electrical contacts cure, the sample is tested with an external measuring system to obtain the temperature coefficient of resistance (TCR) for each on-chip heater. A current flows through the 600 square on-chip heater, and temperature is measured using a K-type thermocouple. From there, a graph of the on-chip heater resistance versus substrate temperature can be plotted as shown in Fig.~\ref{fig2} c. Then, a linear regression fit yields the temperature coefficient of resistance. This TCR value will be used in the temperature calculation of the on-chip heater in the subsequent steps. A wide range of materials has been tested with a focus on silver. We obtain a reasonably constant value for the temperature coefficient of resistance of silver, which is 3.5$\times$10$^{-4}$ K$^{-1}$. This value is quite close to the standard value of bulk material at 3.8$\times$10$^{-4}$ K$^{-1}$ \cite{CRCbook}.

Bi$_2$Te$_3$ thin films are thermally co-evaporated from two separate sources of bismuth and tellurium through an alumina shadow mask with a six square straight bar geometry size 2 $\times$ 12 mm, as shown in Fig.~\ref{fig1} a and b. All materials are of 5N quality. Throughout the evaporation, the ﬂow rates of the two materials are strictly maintained at 0.50 \r{A}/s for bismuth and at 1.25 \r{A}/s for tellurium, which guarantees the stoichiometry for Bi$_2$Te$_3$ \cite{Dang25}. This is a demanding step, as the evaporation rate is cumulatively sensitive to the supplied currents. The substrate temperature is maintained at 474 K to ensure the dominance of the Bi$_2$Te$_3$ phase. Those parameters are all carefully optimized in advance for samples over 100 nm in thickness \cite{Freik, Gonzalez16, Thao}. During thermal evaporation, measurement data are automatically collected via a computer connected with Keithley 2000 multimeters through GPIB/USB ports. The data is collected continuously during the evaporation over time, so each point is only measured once. The uncertainty of each measurement point comes from the multimeter, which is too small and can be ignored. The on-chip heater generates a temperature gradient to measure the Seebeck coefficient. A voltage range of 7-10 V is supplied from a homemade battery box to avoid noises and ground loops. This way, a constant temperature gradient in the range of 1-2 K between the two ends of the bar structure is maintained. The voltage difference measured on these two ends is proportional to the Seebeck coefficient, determined through the TCR of the on-chip heater. 

\begin{figure}[t]
\begin{center}
\includegraphics[width=3in,keepaspectratio]{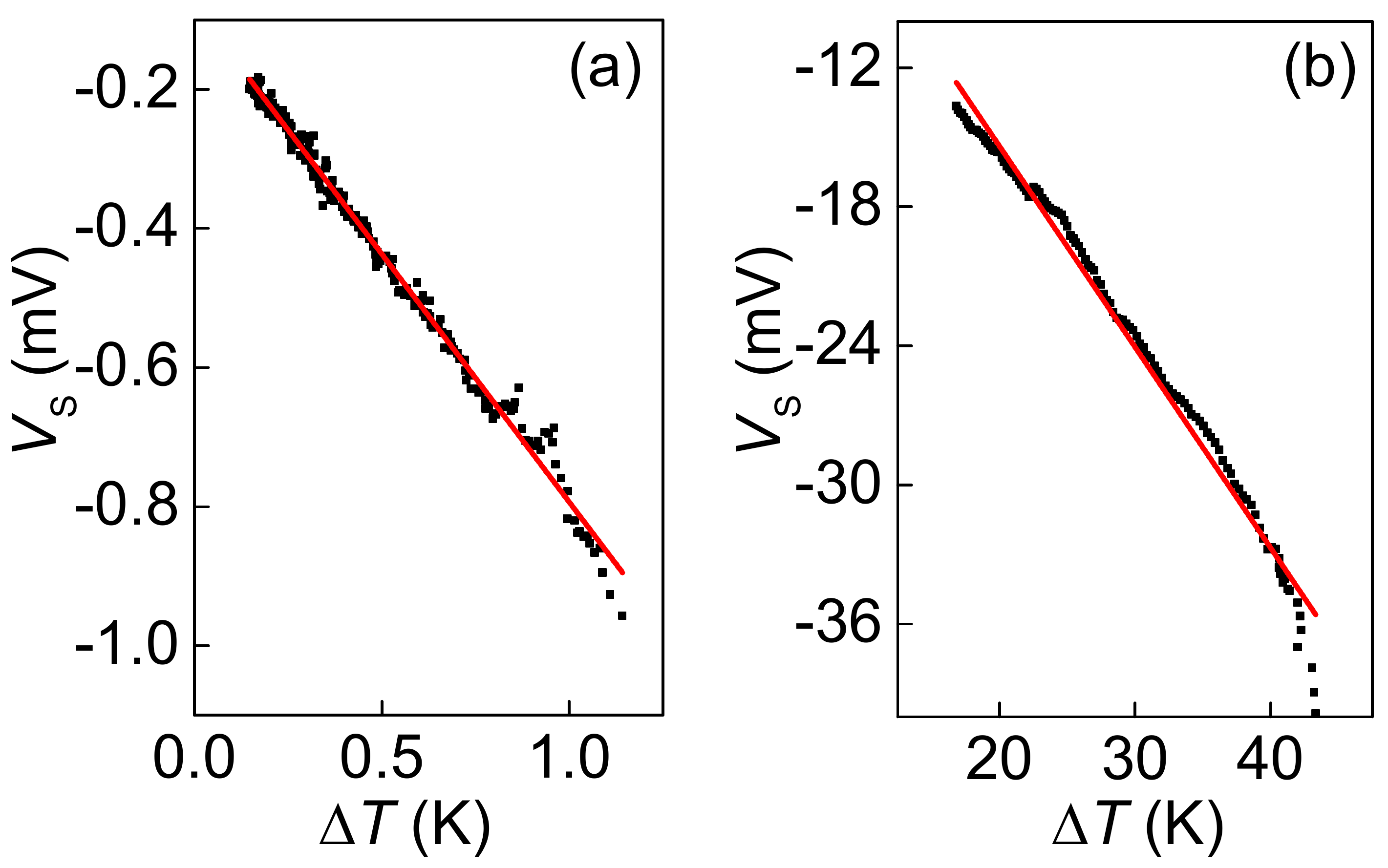}
\caption{\textbf{Comparing the in-situ and ex-situ Seebeck measurement systems: }  A common Bi$_2$Te$_3$ is used to compare the operation of both measurement setups at room temperature. (a) After deposition and inside the vacuum chamber, a battery box is connected to the on-chip heater apparatus that provides a temperature gradient in the range of 1 K, as shown in Fig.~\ref{fig1}. The temperature is interpolated from the measured voltage using its TCR, such as those shown in Fig.~\ref{fig2} c. The slope yields a Seebeck coefficient of 713 $\pm$ 4 $\mu$V/K. (b) Outside the evaporator, the same film is measured in our standard setup as shown in Fig.~\ref{fig2}. Powered by thermoelectric coolers, the temperature difference is much larger. The Seebeck coefficient obtained through linear fit is 868 $\pm$ 10 $\mu$V/K.}
\label{fig3}
\end{center}
\end{figure}

To calibrate the operation of the in-situ Seebeck measurement system, a six square straight bar geometry Bi$_2$Te$_3$ film is deposited at 474 K temperature using our standard process \cite{Thao, Dang25}. Right after the deposition, the heater is set at 301 K. The on-chip heater resistance and the Seebeck voltage of the sample are monitored simultaneously using two multimeters with data logging every 0.1 seconds, as shown in Fig.~\ref{fig1}. The resistance of the on-chip heater represents the temperature of the hot end by fitting it to the temperature coefficient of resistance. We consider the other end of the sample to have a constant temperature throughout the measurement, which is the temperature of the stage. From there, we based on the results of the Seebeck potential difference and the heating structure resistance to calculate the Seebeck coefficient through a linear fit. This data is shown in Fig.~\ref{fig3} a, where the slope gives a Seebeck coefficient of 713 $\pm$ 4 $\mu$V/K. Afterward, this sample is removed from the chamber and re-measured using the ex-situ measurement system, as shown in Fig.~\ref{fig2}. Here, the temperature difference is generated by two commercial thermoelectric coolers and is measured using two K-type thermocouples. The sample is placed on two copper heat sinks to stabilize the temperature generated by the thermoelectric coolers. Thermal conducting paste is used to enhance the thermalization between the heat sink, the cooler, and the Bi$_2$Te$_3$ sample. The temperature difference and the Seebeck potential difference are measured simultaneously, and the data is automatically acquired and then linearly fitted to yield the Seebeck coefficient. Similar to our previous work \cite{Thao, Dang25}, this measurement is performed inside a metallic vacuum chamber for better electrical signal and thermal isolation. The ex-situ measurement result is shown in Fig.~\ref{fig3} b, where the linear slope gives a Seebeck coefficient of 868 $\pm$ 10 $\mu$V/K. 

\begin{table}
  \begin{center}
\begin{tabular}{|c|c|c|c|}\hline
  Material & $d_c$ (nm) & $R_{c,0}$ (k$\Omega$) & $R_{c,10}$ (k$\Omega$)\\\hline \hline
  Bi & 18.6  & 3500  & 250 \\
  Te & 7.2  & 2829  & 1078 \\ 
  Bi$_2$Te$_3$ & 10.1  & 516  & 38 \\
  \hline
\end{tabular}
\end{center}
\caption{Critical thickness and resistance of some materials tested in our setup. $d_c$ is the critical thickness when the film starts to conduct. $R_{c,0}$ is the resistance at the critical thickness, and $R_{c,10}$ is the resistance of the film at thickness $d = d_c$ + 10 nm}
\end{table}

The measurement results of both measurements show that the Seebeck potential $V_S$ linearly depends on the estimated temperature difference at two ends of the sample. Although coming from the same sample, the Seebeck coefficients expressed through the slopes of the graph in Fig.~\ref{fig3} in the two measurements differ by about 10\%. A simple way to look at this discrepancy could come from the difference in the area of the on-chip heater and the contact pad. As seen in Fig. 1, the contact pad is surrounded by the on-chip heater, and its actual temperature might be different from the measured temperature of the heater. We estimate the areas differ by 5/4, which would lead to a mismatch of the two temperatures of about 5/4. This difference matches well to the difference in the Seebeck coefficients obtained from the in-situ and ex-situ setup. However, this discrepancy could associate to more complicated physics. Since the on-chip heater temperature is interpolated from its voltage value, it is the electron temperature of the film. At room temperature, this electron temperature equals the phonon temperature of the metallic heater. However, the interface between the heater and the substrate and the interface between the substrate and the Bi$_2$Te$_3$ film might cause some mismatches due to Kapitza thermal resistance at the boundary. Moreover, thermal convection and substrate thermalization inside the evaporator might be different compared to the ex-situ setup, which might cause a slight change in temperature reading. Finally, the freshly deposited film inside the evaporator might have a different morphology than the film once exposed to the external condition with oxygen and water in the air. The changes in film morphology might lead to changes in its transport and thermoelectricity. 

\begin{figure}[t]
\begin{center}
\includegraphics[width=3in,keepaspectratio]{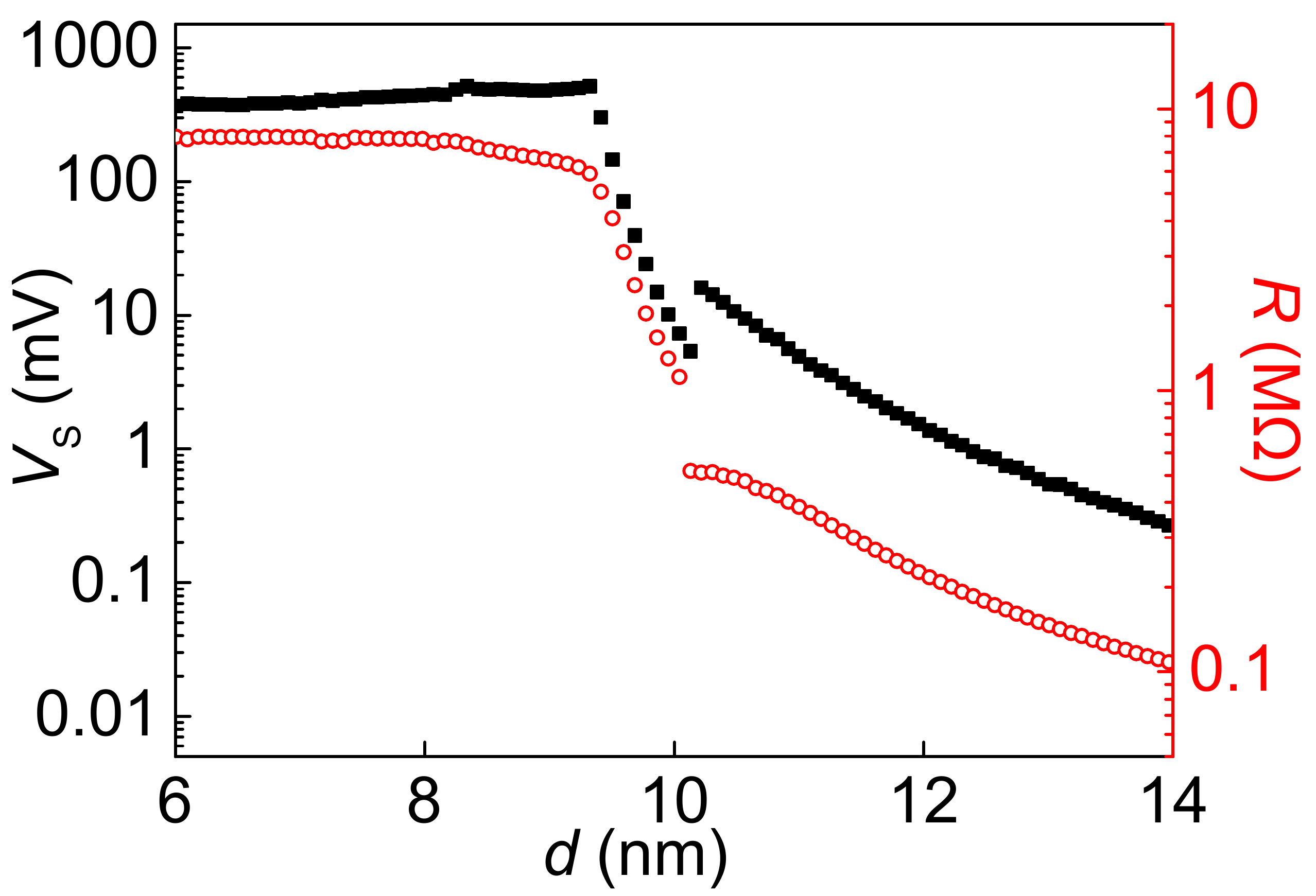}
\caption{\textbf{Performance of the in-situ measurement system:} Seebeck voltage (solid black squares, left axis, in log scale) and resistance (open red circles, right axis, in log scale) of a Bi$_2$Te$_3$ film in the straight bar structure measured during the thermal evaporation as a function of film thickness. In the region of ultra-thin thickness below 10 nm, the thin film has not been formed between the two electrodes. The received signal is in the form of noise. At the critical point, a conducting channel is formed. The film conducts. The Seebeck potential and the resistance of the film start to be measurable. The resistance of the sample changes inversely proportional to the film thickness. The initial Seebeck coefficient gradually decreases with increasing thickness and approaches a stable value agreeing with the bulk value.}
\label{fig4}
\end{center}
\end{figure}

\section{Results and Discussions}

With a calibrated in-situ Seebeck measurement system, we are ready for its demonstration. An experiment is performed where both Seebeck voltage $V_S$ and the film resistance of Bi$_2$Te$_3$ are measured continuously during its evaporation. Here, the evaporation rate for Bi is 0.50 \r{A}/s, and Te is 1.25 \r{A}/s \cite{Thao, Dang25}. The temperature is fixed at 474 K, and the voltage on the on-chip heater is 7 V. At the beginning of the deposition, the film forms as separate islands on the empty Si substrate. These grains are not connected. The measured thickness is the average thickness of these islands. Without connectivity, the film's resistivity is in the range of M$\Omega$, which is the limit of the fixed measurement range on the multimeter. The Seebeck voltage is not defined and shows large fluctuations. Both $V_S$ and $R$ values remain almost constant values as the film grows. This noisy behavior barely changes when the thickness increases. 

At the critical point when at least one conducting channel connects these grains, the film begins to conduct. The resistance $R$ and Seebeck voltage $V_S$ are measurable. In Fig.~\ref{fig4}, the sharp jump shows the moment when the two contact pads are connected. Apparently, it occurs at the same thickness for both $R$ and $V_S$. As shown in Table 1, these critical points differ with different materials. Each material has a different wetting on Si substrates, which lead to different grain size during its evaporation. For each material, we have made multiple measurements to determine the critical point, and they also has slightly different critical points due to random formations of islands on the substrate, the incompatibility uniform in applying silver glue to the contact pad, or the detailed substrate cleanliness. This deviation of the critical point from the same material is typically smaller than 1 nm and can be neglected. These critical point $d_c$ might seems larger than a few atomic layers. Thermal-evaporation at room temperature on generic Si wafer produce poly-crystal films, which lead to a thicker percolation threshold. We associate this observation to an effective critical point \cite{Thao}, where the actual thickness that contributes to conduction might be thinner than the measured value on the crystal oscillators. 

Beyond the critical point, the connection is established. Quantum confinement effect is discernible as both $R$ and $V_S$ sharply decrease with thickness. In about 2 nm, the change in $R$ is about two orders of magnitude, and the change in $V_S$ is three orders of magnitude, as shown in the log scale of the $y$ axis in Fig.~\ref{fig4}. These abrupt changes seem to follow Arrhenius if $R$ and $V_S$ are plotted in the log scale versus $1/d$. We do not explore this effect in detail as it is not the main purpose of this work. As the film grows, the exponential behavior smooths out. $R$ and $V_S$ still reduce with thickness but with a much slower rate. This trend continues until the film approaches a much thicker thickness near the bulk value. The jump in Fig.~\ref{fig4} relates to a glitch in the heating electronics of the evaporator. This observation is entirely consistent with previous results obtained \cite{Thao} when samples are fabricated and measured independently. And once again, the theoretical calculation results based on Boltzmann transport equation \cite{Hicks2D, Bach} is proved. 

Apparently, our in-situ apparatus can resolve thickness to the sub-nanometer range and measure the Seebeck voltage to $\mu$V without amplification. We anticipate that these resolutions are enough to probe the thermoelectricity of a general  material beyond Bi$_2$Te$_3$. More importantly, this setup works at high temperatures and high vacuum, thus being compatible with other types of fabrication recipes. However, our measurement system can still be further improved. The step cover of the ultra-thin film to the silver epoxy depends on the quality of the epoxy and might be a challenge for inexperienced users. We are also limited to a 500 K upper temperature, as dictated by the quality of the Duralco silver epoxy. In the future, this silver epoxy could be replaced by metallic clamps on pre-patterned contact pads. The upper limit for temperature then depends on the thermal evaporator. 

In summary, our design and construction of the in-situ measurement system prove that it works eﬃciently in the ultra-high vacuum regime and at high-temperature conditions beyond 500 K. The on-chip heater generates a sufficient temperature gradient to measure the Seebeck coefficient of an arbitrary sample. This in-situ system can measure the Seebeck coefficient with accurate results but still needs further adjustment to reduce the small discrepancy compared to the ex-situ measurement system. Our perspective is to measure the dependence of the Seebeck coeﬃcient on the thickness and study the quantum conﬁnement eﬀect in this ultra-thin film regime. We are keen to explore the oscillatory nature of the Seebeck effect due to quantum confinement \cite{Rogacheva12, Rogacheva13, Rogacheva14}. Moreover, we expect that this in-situ measurement system would speed up our search for a better not only Bi$_2$Te$_3$ but also many other thermoelectric materials in the quantum regime by reducing the experimenting time to obtain an optimized set of fabricating parameters. It can also be used to study thermoelectricity on other parameters, such as temperature or annealing time.

\emph{Acknowledgement} -- Nguyen Trung Kien was funded by Vingroup JSC and supported by the Master, Ph.D. Scholarship Programme of Vingroup Innovation Foundation (VINIF), Institute of Big Data, code VINIF.2021.TS.057. Samples were fabricated and measured at the Nano and Energy Center, VNU University of Science, Vietnam National University, Hanoi. 

The data that support the findings of this study are available from the corresponding author upon reasonable request.


\begin{thebibliography}{70}
\bibitem{Hicks1D} L. D. Hicks and M. S. Dresselhaus, Thermoelectric ﬁgure of merit of a one- dimensional conductor, Physical review B \textbf{47}, 16631 (1993).
\bibitem{Hicks2D} L. D. Hicks and M. S. Dresselhaus, Effect of quantum well structures on the thermoelectric ﬁgure of merit, Physical review B \textbf{47}, 12727 (1993).
\bibitem{Bach} V. A. Tran, P. A. Tran, H. Q. Nguyen, G. H. Bach, and T. T. Nguyen, Boundary-scattering induced Seebeck coefficient enhancement in thin films within relaxation time approximation, Physica B: condensed matter \textbf{635}, 413800 (2022).
\bibitem{Freik} D. M. Freik, I. K. Yurchyshyn, V. Y. Potyak, and Y. V. Lysiuk, Oscillatory thickness dependences of the Seebeck coefﬁcient in nanostructures based on compounds IV–VI, Journal of materials research \textbf{27}, 1157 (2012).
\bibitem{Ren} J. Mao, Z. Liu, and Z. Ren, Size effect in thermoelectric materials, npj quantum materials \textbf{1}, 16028 (2016).
\bibitem{Kanatzidis} C. J. Vineis, A. Shakouri, A. Majumdar, and M. G. Kanatzidis, Nanostructured thermoelectrics: Big efﬁciency gains from small features, Advanced materials \textbf{22}, 3970 (2010).
\bibitem{Venkatasubramanian} R. Venkatasubramanian, E. Siivola, T. Colpitts, and B. O’Quinn, Thin ﬁlm thermoelectric devices with high room temperature ﬁgures of merit, Nature \textbf{413}, 597 (2001).
\bibitem{Hicks96} L. D. Hicks, T. C. Harman, X. Sun, and M. S. Dresselhaus, Experimental study of the effect of quantum-well structures on the thermoelectric figure of merit, Physical Review B \textbf{53}, 16 (1996).
\bibitem{Thao} T. T. T. Nguyen, L. T. Dang, G. H. Bach, T. H. Dang, K. T. Nguyen, H. T. Pham, T. Nguyen-Tran, T. V. Nguyen, T. T. Nguyen, and H. Q. Nguyen, Enhanced thermoelectricity at the ultra-thin film limit,  Applied physics letters \textbf{117}, 083104 (2020).
\bibitem{Rogacheva9} E. I. Rogacheva, A. V. Budnik, A. Y. Sipatov, O. N. Nashchekina, and M. S. Dresselhaus, Thickness dependent quantum oscillations of transport properties in topological insulator Bi$_2$Te$_3$ thin ﬁlms, Applied physics letters \textbf{106}, 053103 (2015).
\bibitem{Rogacheva10} E.I. Rogacheva, A.V. Budnik, M.V. Dobrotvorskaya, A.G. Fedorov, S.I. Krivonogov, P.V. Mateychenko, O.N. Nashchekina, and A.Yu. Sipatov, Growth and structure of thermally evaporated Bi$_2$Te$_3$ thin films, Thin Solid Films \textbf{612}, 128-134 (2016).
\bibitem{Rogacheva11} E.I. Rogacheva, A.V. Budnik, O.N. Nashchekina, A.V. Meriuts, and M.S. Dresselhaus, Quantum Size Effects in Transport Properties of Bi$_2$Te$_3$ Topological Insulator Thin Films, Journal of electronic materials \textbf{46}, No. 7 (2017).
\bibitem{Rogacheva12} E.I. Rogacheva, S.I. Menshikova, A.Yu Sipatov, and O.N. Nashchekina, Thickness-dependent quantum oscillations of the transport properties in bismuth selenide thin films, Thin solid films \textbf{684}, 31-35 (2019).
\bibitem{Rogacheva13} E. I. Rogacheva, O. N. Nashchekina, A. V. Meriuts, S. G. Lyubchenko, M. S. Dresselhaus, and G. Dresselhaus, Quantum size effects in n-PbTe/p-SnTe/n-PbTe heterostructures, Applied physics letters \textbf{86}, 063103 (2005).
\bibitem{Rogacheva14} E. I. Rogacheva, O. N. Nashchekina, Y. O. Vekhov, M. S. Dresselhaus, and G. Dresselhaus, Oscillations in the thickness dependences of the room- temperature Seebeck coefﬁcient in SnTe thin ﬁlms, Thin solid films \textbf{484}, 433 (2005).
\bibitem{DuongAT} A. T. Duong, V. Q. Nguyen, G. Duvjir, V. T. Duong, S. Kwon, J. Y. Song, J. K. Lee, J. E. Lee, S. Park, T. Min, J. Lee, J. Kim, S. Cho, Achieving ZT= 2.2 with Bi-doped n-type SnSe single crystals. Nature Communications, \textbf{7}, 13713 (2016).
\bibitem{PhanBT} A.T.T. Pham, H.K.T. Ta, Y.R. Liu, M. Aminzare, D.P. Wong, T.H. Nguyen, N.K. Pham, T.B.N. Le, T. Seetawan, H. Ju, S. Cho, Effect of annealing temperature on thermoelectric properties of Ga and In dually doped-ZnO thin films, Journal of Alloys and Compounds, \textbf{747}, 156 (2018).
\bibitem{Gonzalez16} M. Rull-Bravo, A. Moure, J. F. Fernándezb, and M. Martin-Gonzalez, Skutterudites as thermoelectric materials: revisited, RSC advances \textbf{5}, 41653 – 41667 (2015).
\bibitem{Balandin17} V. Goyal, D. Teweldebrhan, and A. Balandin, Mechanically-exfoliated stacks of thin ﬁlms of Bi$_2$Te$_3$ topological insulators with enhanced thermoelectric per- formance, Applied physics letters \textbf{97}, 133117 (2010).
\bibitem{Shinohara18}  M. Goto, M. Sasaki, Y. Xu, T. Zhan, Y. Isoda, and Y. Shinohara, Control of p-type and n-type thermoelectric properties of bismuth telluride thin films by combinatorial sputter coating technology, Applied surface science \textbf{407}, 405 (2017). 
\bibitem{Leu19} P. H. Le, C. N. Liao, C. W. Luo, and J. Leu, Thermoelectric properties of nano- structured bismuth–telluride thin ﬁlms grown using pulsed laser deposition, Journal of alloys and compounds \textbf{615}, 546 (2014).
\bibitem{Jitov20} P. I. Kuznetsov, V. O. Yapaskurt, B. S. Shchamkhalova, V. D. Shcherbakov, G. G. Yakushcheva, V. A. Luzanov, and V. A. Jitov, Growth of Bi2Te3 ﬁlms and other phases of Bi-Te system by MOVPE, Journal of crystal growth \textbf{455}, 122 (2016).
\bibitem{Fedorov21} A. V. Budnik, E. I. Rogacheva, V. I. Pinegin, A. Yu. Sipatov, and A. G. Fedorov, Effect of initial bulk material composition on thermoelectric properties of Bi$_2$Te$_3$ thin ﬁlms, Journal of electronic materials \textbf{42}, 1324 – 1329 (2013). 
\bibitem{Zheng22}  P. Fan, P.-cheng Zhang, G.-xing Liang, F. Li, Y.-xing Chen, J.-ting Luo, X.-hua Zhang, S. Chen, and Z.-hao Zheng, High-performance bismuth telluride thermoelectric thin films fabricated by using the two-step single-source thermal evaporation, Journal of alloys and compounds \textbf{819}, 153027 (2020). 
\bibitem{Min23}  H. Zou, D.M. Rowe, and G. Min, Growth of p- and n-type bismuth telluride thin films by co-evaporation, Journal of crystal growth \textbf{222}, 82 (2001). 
\bibitem{Correia24} L. M. Goncalves, C. Couto, P. Alpuim, A. G. Rolo, F. Volklein, and J. H.Correia, Optimization of thermoelectric properties on Bi2Te3 thin ﬁlms deposited by thermal co-evaporation, Thin Solid Films \textbf{518}, 2816 (2010).
\bibitem{Dang25} L. T. Dang, T. H. Dang, T. T. T. Nguyen, T. T. Nguyen, H. M. Nguyen, T. V. Nguyen, and H. Q. Nguyen, Thermoelectric micro-refrigerator based on bismuth/antimony telluride, Journal of electronic materials \textbf{46}, 3660 (2017).
\bibitem{Oh26} J. H. Kim, J. Y. Choi, J. M. Bae, M. Y. Kim, and T. S. Oh, Thermoelectric characteristics of n-type Bi$_2$Te$_3$ and p-type Sb$_2$Te$_3$ thin ﬁlms prepared by co-evaporation and annealing for thermopile sensor applications, Materials transactions \textbf{54}, 618 (2013).
\bibitem{CRCbook} F. H. DeLand, CRC Handbook of Chemistry and Physics, CRC Press, Inc., ISBN 0849304636, 2386 pp., (1984).


\end{thebibliography}
\end{document}